\documentstyle[12pt,epsf]{article}

\catcode`\@=11
\long\def\@makefntext#1{
\protect\noindent \hbox to 3.2pt {\hskip-.9pt
$^{{\ninerm\@thefnmark}}$\hfil}#1\hfill}                

\def\@makefnmark{\hbox to 0pt{$^{\@thefnmark}$\hss}}  

\def\ps@myheadings{\let\@mkboth\@gobbletwo
\def\@oddhead{\hbox{}
\rightmark\hfil\ninerm\thepage}
\def\@oddfoot{}\def\@evenhead{\ninerm\thepage\hfil
\leftmark\hbox{}}\def\@evenfoot{}
\def\sectionmark##1{}\def\subsectionmark##1{}}

\renewcommand{\thefootnote}{\fnsymbol{footnote}}

\newcounter{sectionc}\newcounter{subsectionc}\newcounter{subsubsectionc}
\renewcommand{\section}[1] {\vspace*{0.6cm}\addtocounter{sectionc}{1}
\setcounter{subsectionc}{0}\setcounter{subsubsectionc}{0}\noindent
        {\normalsize\bf\thesectionc. #1}\par\vspace*{0.4cm}}
\renewcommand{\subsection}[1] {\vspace*{0.6cm}\addtocounter{subsectionc}{1}
        \setcounter{subsubsectionc}{0}\noindent
        {\normalsize\it\thesectionc.\thesubsectionc. #1}\par\vspace*{0.4cm}}
\renewcommand{\subsubsection}[1]
{\vspace*{0.6cm}\addtocounter{subsubsectionc}{1}
        \noindent
{\normalsize\rm\thesectionc.\thesubsectionc.\thesubsubsectionc.
        #1}\par\vspace*{0.4cm}}

\newcounter{appendixc}
\newcounter{subappendixc}[appendixc]
\newcounter{subsubappendixc}[subappendixc]

\renewcommand{\appendix}[1] {\vspace*{0.6cm}
        \refstepcounter{appendixc}
        \setcounter{figure}{0}
        \setcounter{table}{0}
        \setcounter{equation}{0}
        \renewcommand{\thefigure}{\Alph{appendixc}.\arabic{figure}}
        \renewcommand{\thetable}{\Alph{appendixc}.\arabic{table}}
        \renewcommand{\theappendixc}{\Alph{appendixc}}
        \renewcommand{\theequation}{\Alph{appendixc}.\arabic{equation}}
        \noindent{\bf Appendix \theappendixc #1}\par\vspace*{0.4cm}}

\def\abstracts#1{{

\centering{\begin{minipage}{12.2truecm}\footnotesize\baselineskip=12pt\noindent
        \centerline{\footnotesize ABSTRACT}\vspace*{0.3cm}
        \parindent=0pt #1
        \end{minipage}}\par}}

\renewenvironment{thebibliography}[1]
        {\begin{list}{\arabic{enumi}.}
        {\usecounter{enumi}\setlength{\parsep}{0pt}
\setlength{\leftmargin 1.25cm}{\rightmargin 0pt}
         \setlength{\itemsep}{0pt} \settowidth
        {\labelwidth}{#1.}\sloppy}}{\end{list}}

\newcounter{itemlistc}
\newcounter{romanlistc}
\newcounter{alphlistc}
\newcounter{arabiclistc}

\newcommand{\fcaption}[1]{
        \refstepcounter{figure}
        \setbox\@tempboxa = \hbox{\footnotesize Fig.~\thefigure. #1}
        \ifdim \wd\@tempboxa > 6in
           {\begin{center}
        \parbox{6in}{\footnotesize\baselineskip=12pt Fig.~\thefigure. #1}
            \end{center}}
        \else
             {\begin{center}
             {\footnotesize Fig.~\thefigure. #1}
              \end{center}}
        \fi}

\newcommand{\tcaption}[1]{
        \refstepcounter{table}
        \setbox\@tempboxa = \hbox{\footnotesize Table~\thetable. #1}
        \ifdim \wd\@tempboxa > 6in
           {\begin{center}
        \parbox{6in}{\footnotesize\baselineskip=12pt Table~\thetable. #1}
            \end{center}}
        \else
             {\begin{center}
             {\footnotesize Table~\thetable. #1}
              \end{center}}
        \fi}

\def\@citex[#1]#2{\if@filesw\immediate\write\@auxout
        {\string\citation{#2}}\fi
\def\@citea{}\@cite{\@for\@citeb:=#2\do
        {\@citea\def\@citea{,}\@ifundefined
        {b@\@citeb}{{\bf ?}\@warning
        {Citation `\@citeb' on page \thepage \space undefined}}
        {\csname b@\@citeb\endcsname}}}{#1}}

\newif\if@cghi
\def\cite{\@cghitrue\@ifnextchar [{\@tempswatrue
        \@citex}{\@tempswafalse\@citex[]}}
\def\citelow{\@cghifalse\@ifnextchar [{\@tempswatrue
        \@citex}{\@tempswafalse\@citex[]}}
\def\@cite#1#2{{$\null^{#1}$\if@tempswa\typeout
        {IJCGA warning: optional citation argument
        ignored: `#2'} \fi}}

 1
 1
 1

\font\ninerm=cmr9

\def\PRL#1&#2&#3&{\sl Phys. Rev. Lett., \ \bf #1\rm(19#2)#3}
\def\PR#1&#2&#3&{\sl Phys. Rep., \ \bf #1\rm(19#2)#3}
\def\PRD#1&#2&#3&{\sl Phys. Rev., \ \bf #1\rm(19#2)#3}
\def\NP#1&#2&#3&{\sl Nucl. Phys., \ \bf #1\rm(19#2)#3}
\def\PL#1&#2&#3&{\sl Phys. Lett., \ \bf #1\rm(19#2)#3}
\def\ZPC#1&#2&#3&{\sl Z. Phys., \ \bf #1\rm(19#2)#3}
\def\NIM#1&#2&#3&{\sl Nucl. Instr. and Meth., \ \bf #1\rm(19#2)#3}
\def\IJMP#1&#2&#3&{\sl Int. J. Mod. Phys., \ \bf #1\rm(19#2)#3}
\def\RMP#1&#2&#3&{\sl Rev. of Mod. Phys., \ \bf #1\rm(19#2)#3}
\def\INC#1&#2&#3&{\sl Il Nuovo Cimento, \ \bf #1\rm(19#2)#3}
\def\etal{{\it et al.}}
\begin{document}

\centerline {   }
\begin{flushright}
NWU-HEP  95-02
\end{flushright}

\vspace{\fill}

\vskip 1.2 cm
\centerline{\Large\bf Jet production in $\gamma\gamma$ collisions
at TRISTAN}
\baselineskip=22pt
\centerline{\Large\bf with the TOPAZ detector\footnote{
To appear in the proceeding of 10th International Workshop
on Photon-Photon collisions(Photon'95), Sheffield University,
England, April 8-13, 1995}
}
\baselineskip=16pt
\vskip 1 cm
\centerline{\footnotesize Hisaki Hayashii}
\baselineskip=13pt
\centerline{\footnotesize\it Department of physics, Nara women's University}
\baselineskip=12pt
\centerline{\footnotesize\it Nara, 630, Japan}
\centerline{\footnotesize E-mail: hayashii@naras1.kek.ac.jp}
\vskip 1 cm
\baselineskip=13pt
\centerline{\footnotesize\it Representing the TOPAZ Collaboration}

\vskip  10.0 cm
\abstracts{
  Jet production in $\gamma \gamma$ collisions have been studied with
 a jet-cone algorithm using a 233 $pb^{-1}$ data taken by the TOPAZ
 detector at the TRISTAN $e^+e^-$ collider.  The transverse momentum dependence
of the
 jet-cross section  as well as  accompanying activities in the
small-angle
  region show the direct evidence for the resolved photon
 processes. We also present  some properties of the two-jet sample
 with the emphasis on their correlation with the remnant-jet activity.
 }

\newpage
\normalsize\baselineskip=15pt
\setcounter{footnote}{0}
\renewcommand{\thefootnote}{\alph{footnote}}

\section{Introduction}

Among various high-energy collisions, there are large class of
reactions which involve (quasi-) real photons,
such as $\gamma\gamma$,
$\gamma^*\gamma$ and $\gamma p$. The nature of
these reactions are
less well known comparing to our knowledge in $e^+e^-$,
$pp(p\bar{p})$
and lepton-neucleon deep-inelastic(DIS) scattering, because of rather
 complicated dual
(point-like and hadronic)
nature of the (quasi-) real photon.

Recently, the situation has been improved significantly.
Several experimental collaborations have reported
evidence
 for the production of high transverse
 momentum jets in quasi-real $\gamma\gamma$ collisions
in the $e^+e^-$ colliders at
the TRISTAN\cite{TOP93,TOP94,AMY91}and
 LEP\cite{ALEPH93},
and in the $\gamma p$ collisions
at HERA\cite{H192}.
In particular,
we have reported  the evidence of
 substantial activities in the small-angle
region in the sample which has high-$p_{T}$ jets in the central
region\cite{TOP93}. This is a direct evidence of the remnant-jets which are
expected only in the resolved-photon processes\cite{DRE90,AUR94,SAS},
i.e. the processes which
involve
the hard scattering of the partons  inside the photon.
 The observation of the
 remnant-jet has also been reported
in the $\gamma p$ collisions
 at  HERA \cite{H192}.

In this talk, we present
 the updated results of the jet study in
 quasi-real $\gamma\gamma$ collisions with a
larger data sample than the previous publication\cite{TOP93}.
 The inclusive jet-$p_{T}$ distribution have been studied
  down to 1 GeV in order to see individual contribution
more clearly.
The energy distribution
in the small-angle region and the property of two-jet events
have been studied in some detail in order to check the
 consistency with the expectation from the remnant-jets
 in the resolved-photon processes.
It is emphasized that the azimuthal-angle correlation
between two-jets are a sensitive observeable
to the effect of the initial-state radiation
of the incident parton in the photon, as well as
to a possible effect of the multiple scattering,
i.e. events where at least two pairs of partons
scatter whithin
the same $\gamma \gamma$ collision\cite{MUL95}.
 All results presented here are preliminary.

\section{Analysis Method}

\vskip -0.2 cm

The present analysis based on the data collected
at $\sqrt{s}$=58 GeV
corresponding to an integrated luminosity of 233
$\rm{pb}^{-1}$.

 The
hadronic events in quasi-real $\gamma\gamma$
collisions are selected with the criteria:
(1)
 the minimum number of charged tracks must be 4 and the net charge
 should be $\leq 2$, where the charged particles should have
 $p_{t} > 0.15$ GeV
 and $\vert \cos\theta \vert <0.83$.
(2)
 the visible energy $(E_{vis})$ of the event should be
 $E_{vis} \leq 35 \rm{~GeV}$,
(3)
  the invariant mass of
an event($W_{vis}$) should be $2 < W_{vis} < 25$ GeV,
(4)
 the maximum energy($E_{max}$) of
an electromagnetic energy cluster in the region
 $|\cos\theta| < 0.9984(\theta_{min}=3.2^{\circ})$
is required to be
 $\leq 0.25 E_{beam}$.
(5)
 beam-gas events are rejected by
requiring  the event vertex position to be
$r_{vtx}<3$ cm and $z_{vtx}<3$ cm.

In the selection (2) and (3),
both charged tracks and the clusters in
 the region $\vert \cos\theta \vert \leq 0.83$ are
used to evaluate $E_{vis}$ and
 $W_{vis}$.
The cut on  $W_{vis}$ has been changed
from  15 GeV to 25 GeV from the previous
analysis in order to study the event in
wider kinematical region.
The selection (4) defines our anti-tag condition
 corresponding to the maximum photon virtuality of
  $P^{2}_{max}= 2.6 ~\rm{GeV}^{2}$.
\begin{figure}
\vskip 7cm
\includegraphics{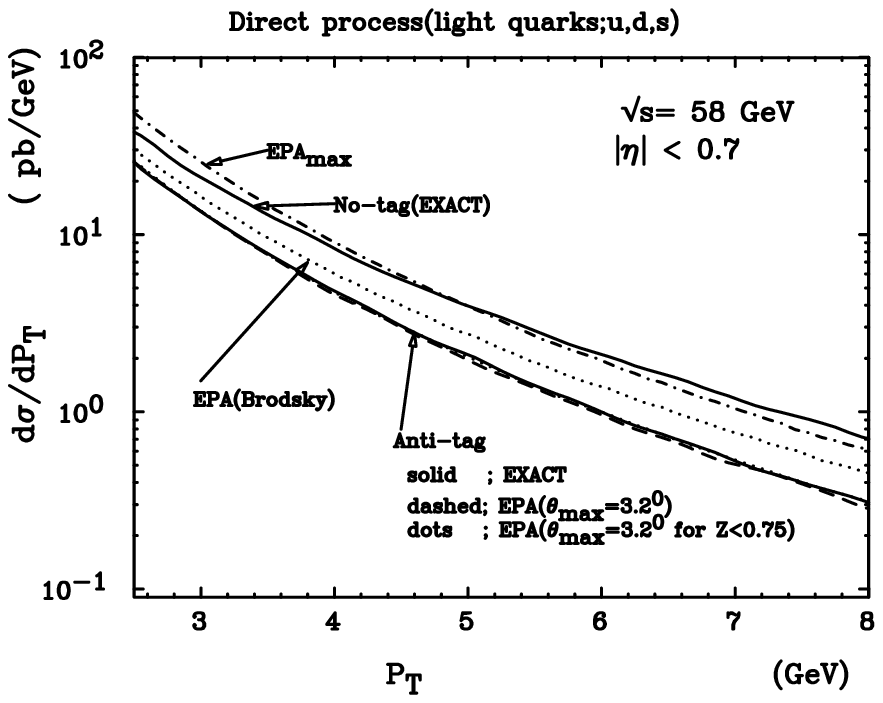}
\vskip -0.9 cm
\caption{
            Comparison of the cross section between EPA approximation and exact
            matrix-element calculation
          in the direct-process.
 The meaning of each curve is explained
in ref.11. }
\label{fig:epa}
\end{figure}
\begin{figure}
\vskip 7cm
\includegraphics{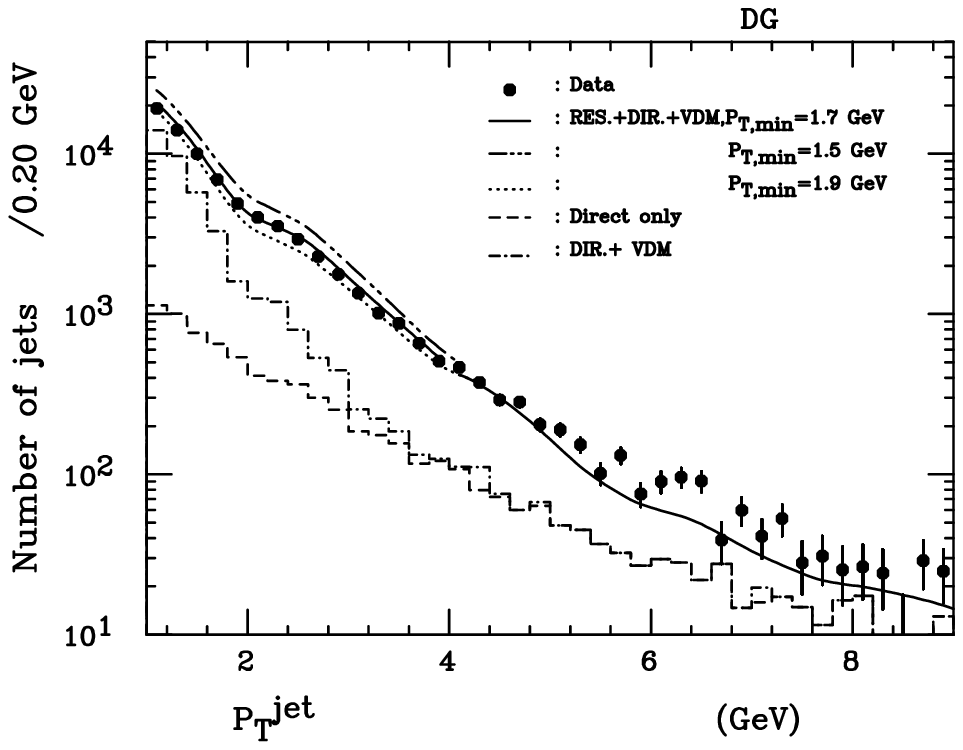}
\vskip -1.0 cm
\caption{
    Inclusive jet distribution as a function of jet transverse
   momentum, $p_{T}(jet)$, for $|\eta_{J}| \leq 0.7$.
  The error bars in data(solid circle) are only statistical ones.
   The solid line a prediction
   of the direct + resolved(DG) + VDM processes with
   $p_{T,min}=1.70$~GeV.
  The dot and two-dots-dashed lines
   correspond to $p_{T,min}=1.5$ and 1.9 ~GeV,respectively.
   The contribution of the direct and the direct+VDM processes are shown
    by dashed and dot-dashed histograms, respectively.
   }
\label{fig:onept}
\end{figure}

 With these criteria and after subtracting background, 112,000
 anti-tag two-photon events are selected in
total.
 The main backgrounds come from beam-gas events in low mass
 region and annihilation events
($e^+e^-\rightarrow q \bar{q} \gamma$) in
 high mass region.
  In all figures
 presented here the backgrounds are subtracted on a bin-by-bin basis.

\vskip 0.5 cm

 The cross section of
 the direct process
($\gamma\gamma \rightarrow q\bar{q}$)
and resolved processes in
$e^+e^-$ collisions are given as a product of
the photon-flux factor, parton density inside the photon,
the subprocess cross section.
 The parton density for the direct process is
given by the delta function
$\delta(1-x)$.
Since the detail description of the generator-program
used in this analysis are
given in
ref.10,
 here we only mention a
few special comments.

The photon-flux factor is given by the formula
of equivalent-photon approximation (
EPA ) :
\begin{eqnarray}
 f_{\gamma/e}(z)& = & \frac{\alpha}{2\pi z} \bigl[ 1+ (1-z)^{2} \bigr]
           \ln \frac{P_{max}^{2} ( 1-z)}{ m_{e}^{2} z^{2} }
           - \frac{\alpha}{\pi} \frac{1-z}{z},               \\
  \rm{with} &  P_{max}^{2} & = \min( P^{2}_{max,kin}, Q^{2}_{eff})
\label{eqn:photon}
\end{eqnarray}
where $Q^{2}_{eff}= p_{T}^{2}$ for light(u,d,s) quarks and
$m_{c}^{2}+ p_{T}^{2}$ for charm quark. $P^{2}_{max,kin}$ is
a maximum photon virtuality determined by the experimental anti-tag
condition: $P^{2}_{max,kin} = 2 E_{beam}(1-z)(1-\cos\theta_{tag})$.
Our  condition is $\theta_{max}=
3.2^{\circ}$ for $z (\equiv E_{\gamma}/E_{beam}) < 0.75$.
To check the precision of this formula, we have compared the
cross section of the direct process obtained from the exact
matrix-element
 calculation\cite{KURO88} and the EPA given in (1).
In our anti-tag situation,
 both results agree in 1 \% level
as shown in
Fig.\ref{fig:epa}.\cite{FRI93}

In the TRISTAN energy region, the range of invariant-mass of $\gamma\gamma$
system($W_{\gamma\gamma}$) for resolved processeses is
between 5 - 50 GeV having a peak at about 20 GeV. This value  is high enough
for perturbative-QCD calculation.
The region of the Bjorken-$x$  is covered down to about 0.1
in the jet analysis presented here.

The Monte-Carlo program is based on the leading-order(LO)
formula, while the
next-leading-order(NLO) calculation is also available
 by Aurenche \etal\cite{AUR94} in mass-less formula.
This NLO results show that,
as far as the energy scale($\mu^{2}$)
appears in $\alpha_{s}$ and the parton-density in the photon
is chosen to the square of the parton-$p_{T}$ in the LO calculation,
 the correction of NLO is small  at high-$p_{T}$
region above 5 GeV.
\cite{AURPRV}

In the current program, we included
 the contribution from
the charm contents in the photon  in the cross section of
resolved processes
 if the center of mass in the subsystem
$\sqrt{\hat{s}}$ is greater than 5 GeV.

\begin{figure}[bh]
\vskip  2.5 cm
\vskip 6.5cm
\includegraphics{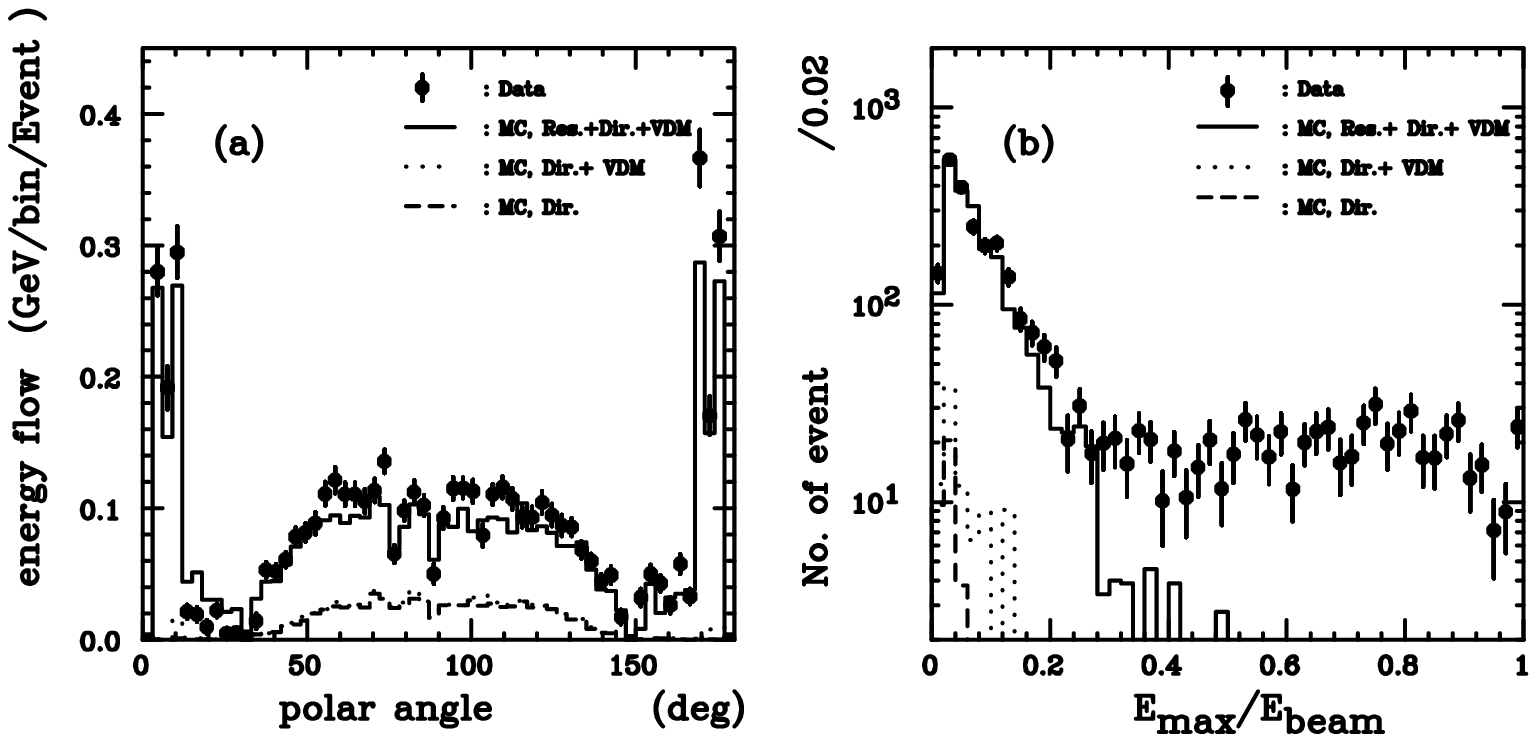}
\vskip -2.6 cm
\caption{
   (a)    Energy flow distribution for the central two-jet sample
       with $\vert \eta_{J} \vert \leq $ 0.7 and
       $p_{T}^{jet} \geq 2.0$ GeV.
     (b)    Maximum
cluster-energy distribution measured by the
       calorimeters in the polar angular region of
         $0.972 < \vert \cos\theta \vert < 0.9984$ for
 the two-jet sample
  without anti-tag condition.
         }
\label{fig:eflcmb}
\end{figure}

\vskip -0.2 cm
\section{Results}
\vskip  -0.5 cm

\subsection{Inclusive Jet Distribution}

 In order to identify jets in the twophoton hadronic events,
   we use a jet-cone
algorithm
 and define
  a jet  as a cluster comprising of
particles inside a circle in the pseudorapidity
$(\eta=-\ln\tan(\theta/2))$- azimuthal angle($\phi$) plane as
$
   \sqrt{ (\triangle\eta^{2} + \triangle\phi^{2})} < 1.
$
Table-1 summarizes
the number of jets and the number of two-jet events selected
for various values of the
cuts in jet-$p_{T}$ and jet-rapidity($\eta_{J}$).
\vskip -0.5cm
\begin{table}[h]
\caption{}
\label{table:jet}
\vskip 0.1cm
\hskip 3 cm
\begin{tabular}{lcccc} \hline
   & $p_{T}$(GeV)  & $\vert \eta_{J} \vert$ &   Data  &       \\ \hline
 No. of jets & $\geq$ 1.0   &  $\leq$ 0.7     &   78637 &  jets  \\
             & $\geq$ 2.0   &  $\leq$ 0.7     &   22198 &  jets  \\
 No. of two-jet event & $\geq$ 2.0  & $\leq$ 0.7 & 2801 &  events \\
                      & $\geq$ 2.0  & $\leq$ 1.0 & 4592 &  events  \\
\hline
\end{tabular}
\vskip -0.2cm
\end{table}

The jet-$p_{T}$ distribution  in the region
$\vert \eta_{J} \vert \leq 0.7$
is shown in Fig.\ref{fig:onept},
 together with the expectation of the VDM(Vector-meson Dominance Model),
 direct and
 resolved processes.
 It is worth  noting  that the slope of data is different
 in the region below and above about
   $p_{T} = 2.0 $~GeV. Below $p_{T}< 2.0$ GeV\cite{VDMJET},
 the VDM components
   are dominated and the above the perturbative components
  of direct and resolved processes
   play an important role.

 The resolved processes are definitely
necessary to reproduce the data typically in the region
 $3 < p_{t}(jet) < 7$ GeV.
 The DG\cite{DG85} parametrization of the parton density of
 the photon is used in this figure. However,
 the data can fit  well  also by the
  LAC1\cite{LAC91} parametrization if the value
 of $p_{T,min}$, a free cut-off parameter  introduced
 in the theory to ensure the
 perturbative-QCD calculation, is optimized.
 The optimum value of $p_{T,min}$
 is found to be
 1.7 GeV for DG and 2.2 GeV for LAC1.
 These optimum values change by 0.05 GeV if the cross section of
 the VDM component\cite{VDM}
 is changed by 30 \%.(See ref.2
 for more detail discussion.).

\subsection{two-jet and remnant-jet }

In the previous paper\cite{TOP93} we  reported
the  evidence for the
remnant-jet activity in small-angle region\cite{TOP93}.
In that analysis, we selected the sample with
at least  one-jet in  central   region,
because of the limited statistics.
 Fig.\ref{fig:eflcmb}(a) show the  distribution of the
particle energy flow
  for the sample which has  two high-$p_{T}$
  jets  in the central region, where these two-jets
 are selected with
 a criteria of
 $p_{T}^{jet} \geq 2.0$ GeV and $\eta_{J} \leq $ 0.7.
 Clearly large
 activities are observed in the small-angle region
 ($0.972 < \vert \cos\theta \vert < 0.9984$).
  The forward calorimeter(FCL) covers in this region.

  It is very rare (less than 1.5\%) that the
direct (or VDM) processes leave some deposit of energies in this
 region if two high-$p_{T}$ jets are required in the central.
The contamination of the beam-gas induced backgrounds
is estimated from the random-trigger sample, and found
to be small (about 10\%), thanks to an extensive
masking system of the detector\cite{MASK93}.

\begin{figure}[hb]
\vskip  1.2 cm
\vskip 7cm
\includegraphics{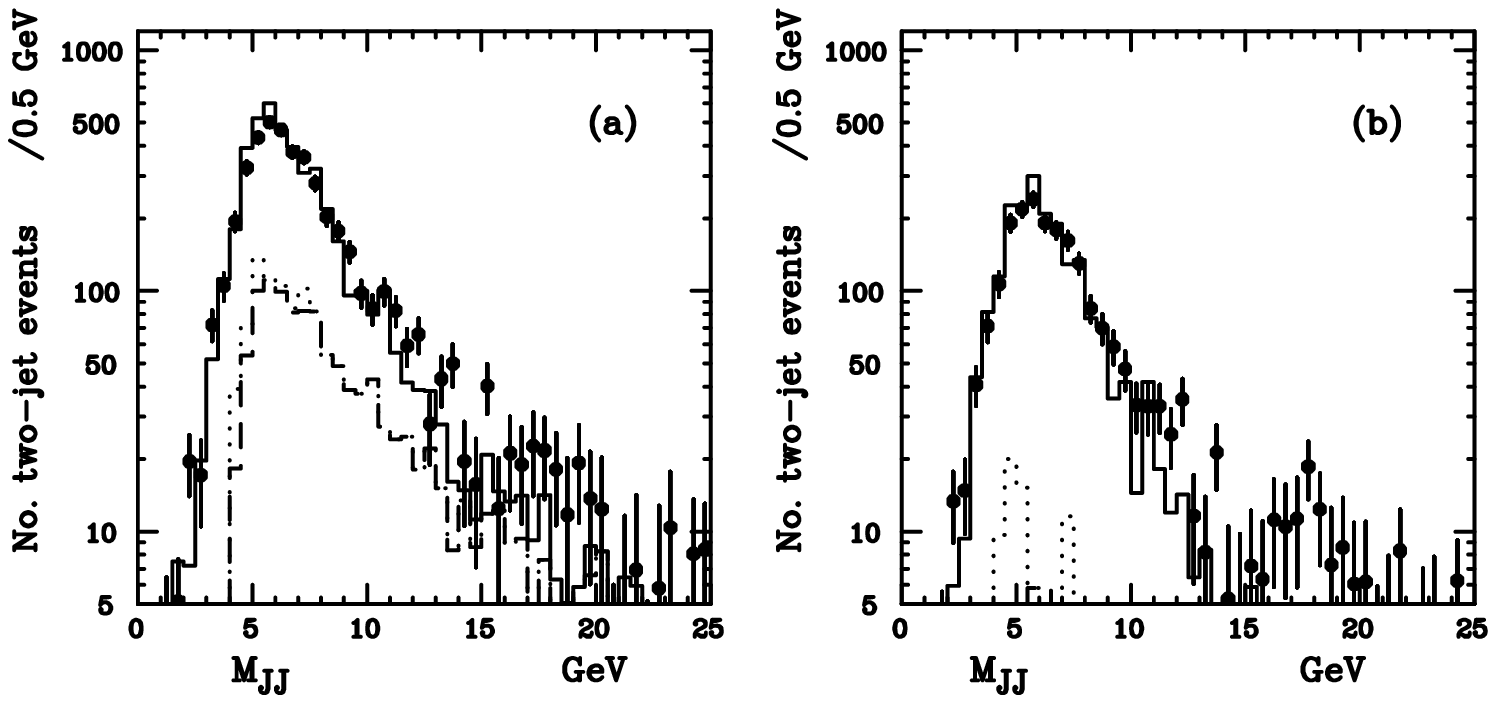}
\vskip -2.0 cm
\caption{
        $\rm{ M_{JJ}}$ distribution for (a) all two-jet,
                            (b) remnant-jet tagged sample.
    The meaning of the histograms is the same as the one given in  Fig.3.}
\label{fig:twomjj}
\end{figure}

Another possible background might come from
 the scattered electron
in deep-inelastic $e\gamma$ scattering.
To study this effect,
no-tag event sample is selected
 without requiring the selection (4) of
 the anti-tag condition.
The largest cluster-energy($E_{max}$)
in  FCL is plotted in
 Fig.\ref{fig:eflcmb}(b) for the no-tag two-jet sample.
In the figure,
  both effects of
the remnant-jets and the scattered electrons
 of the DIS $e\gamma$ scattering are seen clearly
\footnote[1]{The first results from the study of the
jet production in the deep-inelastic
$e\gamma$ is reported in
ref.19. }.
The effect of the scattered electron in the region
 $E_{max}^{FCL} ~<~ 0.25 E_{beam}$,
is less than a few \%.
In this low-energy region,
the Monte-Carlo simulation of resolved processes which
 includes the remnant-jets  explain the
 data  quit well.
\begin{figure}
\vskip  1.2 cm
\vskip 7cm
\includegraphics{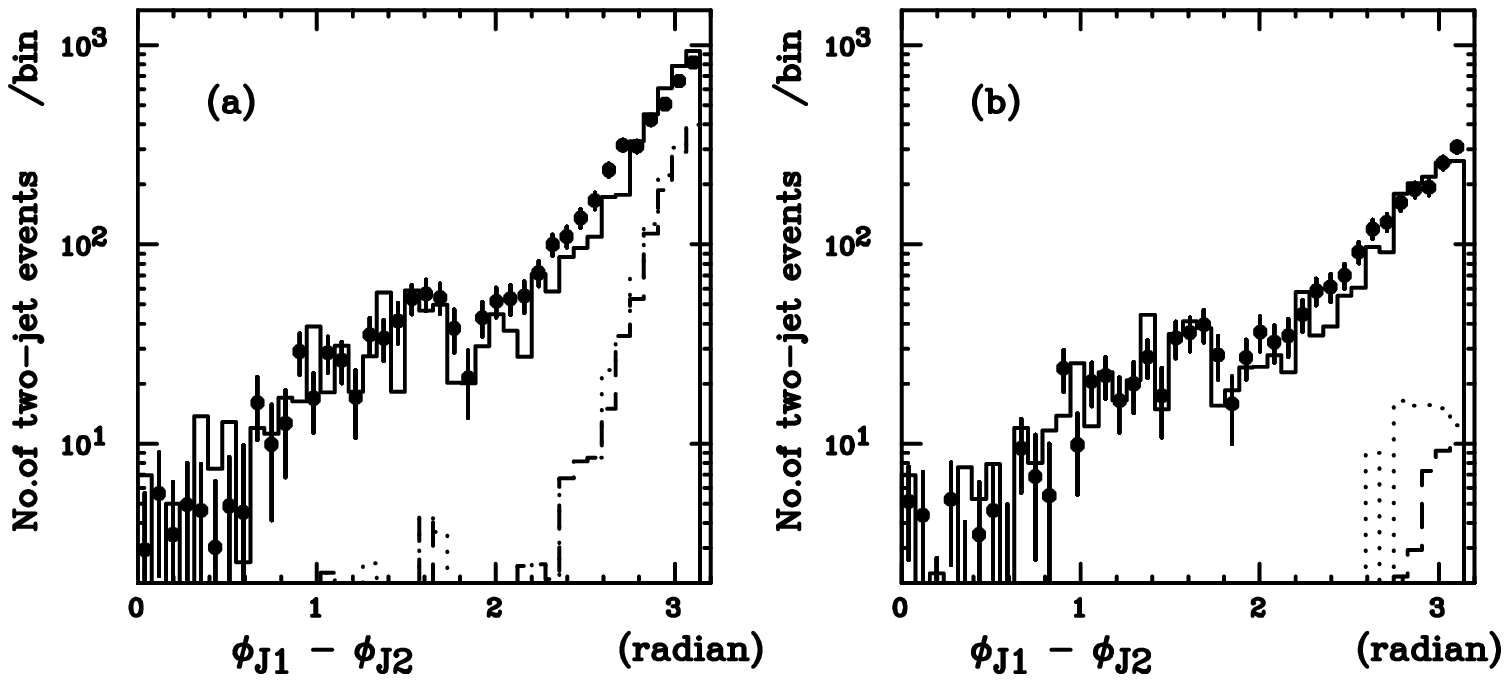}
\vskip -2.0 cm
\caption{ $\triangle\Phi$ distribution for (a) all two-jet,(b)
remnant-jet tagged sample.
   The meaning of the histograms is  the same with
 the one  given in Fig.3.}
\label{fig:twophi}
\end{figure}

 For the detail study of the
  high-$p_{T}$ two-jet events, we require that
 the jet-$p_{T}$ is  greater than 2.0 GeV and
 the
rapidity of both jets less than 1.0.
we also define a sub-sample of two-jets  with a condition:
$ 0.5 \rm{GeV} \geq E_{max}^{FCL} \geq 0.25 E_{beam}$
in order to  enhance the contribution from the  resolved processes
 and
 call them as  remnant-jet-tag events.
With these criteria,
 2172 remnant-tagged events are selected out of the total
  two-jet sample of 4582 events.
  The purity of the resolved process
 in the sub-sample is
 estimated by the Monte Carlo
simulation to be 97 \%
while the tagging efficiency of the 1- and 2-resolved processes
is found to be
 60\%.

  The invariant mass of the two-jet system
  is shown in
Fig.\ref{fig:twomjj}(a) for total and  (b)
 for the remnant-jet-tag  events, respectively.
 The distribution of
  the  opening angle between two-jets in the azimuthal plane
  ($\triangle\phi = \phi_{J1}- \phi_{J2}$)
  are  shown in Fig.\ref{fig:twophi}(a)  also for total and
  (b) remnant-jet-tag  events, respectively.
  We
  note that the substantial part of the two-jet sample
 is almost back-to-back(Fig.\ref{fig:twophi}) as expected.
 As can be seen from these figures,
 the
 LO Monte-Carlo  simulation reproduces
 the overall feature of the data
 in shape and its normalization
 both for total and remnant-jet tagged  sample.
 However,  some deviations are observed in the high-$M_{JJ}$
 region greater than 15 GeV and in the region of
 $ 2 \geq \triangle\phi \geq 2.6$~ radian.
 The latter point would be understood
 as an effect of the initial-state radiation
 of the incident parton
 in the hard-sub-processes which is not included
 in our Monte-Carlo program.
 Further analysis  are in progress.

\vskip 0.3 cm

 In summary
 we have studied the  jet production in the two-photon
 reaction at $\sqrt{s}=58$~ GeV for the integrated luminosity of
 233$\rm{pb}^{-1}$ with the anti-tag condition.
The existence
 of the resolved process in $\gamma\gamma$ collisions
is confirmed from the jet rate and the direct
 detection of the energy-clusters coming from the remnant-jets.
 The rate of the remnant-jet is also consistent with
 the Monte-Carlo expectations.

\vskip -0.3cm
\section{Acknowledgement}
\vskip -0.3cm

The author acknowledges to Drs.  P.~Aurenche, M.~Drees,
R.M.~Godbole, J.Ph.~G\"uillet, K.~Hagiwara,
L.~L\"onnblad,
 G.A.~Schuler, M.~Seymour,
 and I.~Tyapkin for variable
conversations about the analysis.


\begin{thebibliography}{99}
\bibitem{TOP93}
 TOPAZ Collab., H.~Hayashii \etal, \PL B314&93&149&, see also
 ref.2.
\bibitem{TOP94}
  T.~Tauchi, Proceeding of XXIXth Recontres de Moriond
  ``QCD and High Energy Interactions'', Meribel, France, March 19-26, 1994.
\bibitem{AMY91}
     AMY     Collab., R.~ Tanaka \etal, \PL B277&92&215&,
     AMY Collab., B.J.~Kim \etal, \PL B325&94&248&.
\bibitem{ALEPH93}
     ALEPH Collab., D.~Buskulic \etal, \PL B313&93&509&;  \\
     DELPHI Collab., P.~Abreu \etal   \ZPC C62&94&357&.
%
%
\bibitem{H192}
     H1    Collab., T.~ Ahmed \etal, \PL B297&92&205&;    \\
     ZEUS    Collab., M.~Derrick \etal, \PL B297&92&404&.
%
%
\bibitem{DRE90}
    M.~Drees and R.M.~Godbole,           \NP B339&90&355&.
%
%
\bibitem{AUR94}
P.~Aurenche, J.~Ph.~Guillet, M.~Fontannaz, Y.~Shimizu,
 J.~Fujimoto and K.~Kato,
Prog. Theor. Physics {\bf 92}(1994)175, KEK preprint-93-180.
\bibitem{SAS}
  G.A.~Schuler and T.~ Sj\"ostrand, CERN-TH/95-62
\bibitem{MUL95}
M.~Drees, in these proceedings.
\bibitem{MIY95}
 A.~Miyamoto and H.~Hayashii, KEK preprint-94-204.
\bibitem{EPACUR}
  The curve of $\rm{EPA_{max}}$ is the  cross section
  obtained by using the EPA formula(1) with
  $\theta_{max}=180^{\circ}$;
  EPA(Brodsky) is the  one obtained by using another
  EPA-formula given in
  ref.6.
  No-tag(EXACT) is an exact-matrix element calculation
  without any anti-tag condition. In the cases of Anti-tag,
  the dots curve
  (EPA($\theta_{max}=3.2^{\circ}$) for $z<0.75$)  is
   overraped with the dashed curve
  (EPA($\theta_{max}=3.2^{\circ}$)).
\bibitem{AURPRV}
 private communication with P.~Aurenche.
\bibitem{KURO88}
 M.~Kuroda, Meiji Gakuin University(Tokyo) Res. J. {\bf 424}(1988)27.
\bibitem{FRI93}
 S.~Frixione \etal, \PL B319&93&339&,
 K.~Hagiwara \etal, \NP B365&91&544&.
\bibitem{DG85}
    M.~Drees and K.~Grassie, \ZPC C28&85&451&.
%
%
\bibitem{LAC91}
   H.~Abramowicz, K.~Charchula and A.~Levy, \PL B269&91&458&.
\bibitem{VDM}
 For the total $\gamma\gamma$
 cross section due to VDM, we
 used $\sigma_{\gamma\gamma}^{VDM}(W) = (A + B/W)$ nb with
 A=240 and B=270. The final state of
the hadronic system is simulated
 assuming the $p_{t}$ spectra has an exponential form of
 $d\sigma/dp_{t}^{2} \sim e^{-4p_{t}^{2}}$.
\bibitem{MASK93}
             H.~Kichimi \etal, \NIM A334&93&367&.
%
%
\bibitem{TOPF2}
    TOPAZ Collab., K.~Muramatsu \etal, \PL B332&94&477&.
%
%
\bibitem{VDMJET}
The event shape is rather spherical in the $p_{T}$ region less than
about 2 GeV although we applied a jet-cone algorithm
to identify a jet.
While, above 3-4 GeV, clear jet-like signature is seen as
shown in ref.1. This signature can be seen more
easily in the $\eta-\phi$ plane than $\theta-\phi$  plane.
\end{thebibliography}
\end{document}